\documentclass[twocolumn,showpacs,preprintnumbers,amsmath,amssymb]{revtex4}

\usepackage{graphicx,dcolumn,bm}
\usepackage{epsfig,bbold}

\newcommand{\beq}{\begin{equation}}
\newcommand{\eeq}{\end{equation}}
\newcommand{\bea}{\begin{array}}
\newcommand{\eea}{\end{array}}
\newcommand{\bec}{\begin{center}}
\newcommand{\eec}{\end{center}}
\newcommand{\bei}{\begin{itemize}}
\newcommand{\eei}{\end{itemize}}

\begin{document}

\preprint{SACLAY-T07/110, UCRHEP-T435}

\title{Common origin of $\theta_{13}$ and $\Delta m^2_{12}$ in a model of
neutrino mass\\ with quaternion symmetry}

\author{Michele Frigerio$^1$ and Ernest Ma$^2$}
\affiliation{
$^1$ Service de Physique Th\'{e}orique, CEA-Saclay, 
91191 Gif-sur-Yvette Cedex, France
\\
$^2$ Department of Physics and Astronomy, University of California, Riverside, 
California 92521, USA
}

\date{August 6, 2007}

\begin{abstract}
The smallness of the $1-3$ lepton mixing angle $\theta_{13}$ and of the 
neutrino mass-squared-difference ratio $\Delta m^2_{12}/\Delta m^2_{23}$ 
can be understood as the departure from a common limit where they both 
vanish.  We discuss in general the conditions for realizing the mass 
degeneracy of a pair of neutrinos and show that the vanishing of a CP 
violating phase is needed.  We find that the discrete quaternion group $Q$ 
of eight elements is the simplest family symmetry which correlates the 
smallness of $\Delta m^2_{12}$ to the value of $\theta_{13}$.  In 
such a model we predict 
$0.12\lesssim \sin\theta_{13} \lesssim 0.2$ if the ordering of the 
neutrino mass spectrum is normal, and $\sin\theta_{13}\lesssim 0.12$ 
if it is inverted.
\end{abstract}


\maketitle


\noindent{\bf Introduction.~} 
Contrary to quarks and charged leptons, the 
three neutrinos are known to have a mass spectrum with a weak hierarchy, 
possibly quasi-degenerate. In particular, the two mass eigenstates $\nu_1$ 
and $\nu_2$ in the ``solar pair'' are very close in mass in the case of 
inverted ordering and also in the case of normal ordering, as long as the 
yet unknown absolute mass scale is larger than $\sim 0.02$ eV. Indeed, a 
global fit of neutrino oscillation data gives
\beq\bea{l}
\Delta m^2_{12}\equiv m_2^2-m_1^2 = (7.9^{+1.0}_{-0.8}) \cdot 10^{-5} 
{\rm eV}^2~,\\
\Delta m^2_{23}\equiv m_3^2-m_2^2 = \pm(2.6\pm 0.6) \cdot 
10^{-3} {\rm eV}^2~,\\
\sin^2\theta_{12}=0.30^{+0.10}_{-0.06} ~,\\
\sin^2\theta_{23}=0.50^{+0.18}_{-0.16} ~,\\
\sin^2\theta_{13}\lesssim 0.040 ~,
\label{mix}
\eea\eeq
where we took the best fit values and the $3\sigma$ intervals from the
last update of Ref.\cite{MSTV} (v5).

The smallness of the $1-2$ mass splitting compared to the $2-3$ 
``atmospheric'' splitting may be explained as the departure from a symmetric 
limit where $1-2$ mass degeneracy holds.  Other small parameters of the lepton 
flavor sector may also be interpreted as deviations from the same limit, 
such as the $1-3$ mixing angle, the deviation of the $1-2$ and possibly 
$2-3$ angles from the maximal value $\pi/4$, and the mass ratios ($m_e/m_\mu$, 
$m_\mu/m_\tau$) of charged leptons. 

In this paper we analyze the neutrino mass structures corresponding to a 
quasi-degene\-rate pair of states and the phenomenological correlations with 
other lepton flavor parameters, in particular a non-vanishing $1-3$ mixing,
whose measure is the objective of an extensive experimental program \cite{glo}.
We also search for the simplest flavor symmetries which can be used to realize 
such a mass degeneracy. The discrete quaternion group of eight elements $Q$  
is identified as the most suitable for this purpose and a complete model 
is constructed.

The group $Q$ (sometimes called $Q_8$ or $Q_4$) was introduced in \cite{q8} 
to build a model of quark and lepton masses and mixing.  Discrete subgroups 
of quaternions with unit norm, i.e. $SU(2)$, 
were already used in \cite{Getal} to suppress 
the neutrino mass while allowing for large neutrino magnetic moments.  They 
were also discussed as flavor symmetries in a series of papers by Frampton and 
collaborators \cite{Fetal}.  A specific model based on the quaternion group 
of 12 elements $Q_6$ was proposed  as well \cite{baku}.  Models were built 
\cite{ACL} using the binary tetrahedral group $T'$ (double covering of the 
tetrahedral group $A_4$), which is also a discrete subgroup of quaternions 
with 24 elements.  The group $T'$ was recently employed to accommodate 
tri-bi-maximal mixing \cite{Tp}.  Note that $A_4$ is not a 
subgroup of $T'$, but $Q$ is.  Geometrically, $SU(2)$ is isomorphic to the
hypersphere in four dimensions, the 8 elements of $Q$ form 
the 8 vertices of the perfect hyperoctahedron (dual of the hypercube), 
whereas the 24 elements of $T'$ form the 24 vertices 
of the hyperdiamond (which is self-dual).
\\


\noindent {\bf Mass matrix of 2 degenerate neutrinos.~} 
Let us begin 
considering the Majorana mass matrix $m_\nu$ for two neutrino states.
Under the requirement of mass degeneracy ($m_1=m_2=m$), $m_\nu$ can be 
written in full generality as
\beq
D_\phi m_\nu D_\phi =  \left(\bea{cc} a & b \\ b & -a \eea\right) =
\left(\bea{cc} \cos 2\theta & \sin 2\theta \\ \sin 2\theta & -\cos 2\theta 
\eea\right) m ~,
\label{mag}\eeq
where $D_\phi=diag(e^{i\phi_1},e^{i\phi_2})$ is a diagonal matrix of phases, 
$a$ and $b$ are real and positive, $m=\sqrt{a^2+b^2}$ and $\tan2\theta = b/a$. 
In terms of the $m_\nu$ matrix elements, the requirement of mass degeneracy 
is equivalent to 2 conditions, $|m_{11}|=|m_{22}|$ and 
$\arg(m_{11}m_{22}/m^2_{12})=\pi$.

The simplest cases $b=0$ or $a=0$ have often been discussed, while the general 
case was studied in just a few interesting papers, as a prototypical example 
of a pseudo-Dirac neutrino mass matrix \cite{wolf,petcov,JR}.  Here we analyze 
in detail how this matrix structure relates to the physical observables, 
with special attention to the effects of possible CP violating phases.

The diagonalization of the matrix in eq.(\ref{mag}) presents some subtleties, 
which turn out to be important to understand the effect of small perturbations 
responsible for $m_1\neq m_2$. Notice first that the two neutrino masses 
$m_{1,2}$, and the moduli of the mass matrix elements, $a$ and $b$, are 
physically well-defined quantities (they can be measured, at least in 
principle). This determines uniquely the parameter $\theta$, which one is 
tempted to identify with a physical mixing angle $\theta_{12}$ between 
$\nu_1$ and $\nu_2$.  However, in the degenerate limit there is no mixing 
angle responsible for neutrino oscillations: $m_\nu m_\nu^\dag=diag(m^2,m^2)$. 
To settle this apparent contradiction let us rewrite $m_\nu$ in the standard 
parameterization, 
\beq\bea{c}
m_\nu = U^* m_\nu^{diag} U^\dag \\ \\
= D'_\phi 
\left(\bea{cc} c_{12} & s_{12} \\ 
-s_{12} & c_{12}\eea\right) 
\left(\bea{cc} m e^{-2i\rho} & 0 \\ 0 & m \eea\right)
\left(\bea{cc} c_{12} & -s_{12} \\
s_{12} & c_{12}\eea\right) D'_\phi ~, 
\eea\eeq 
where $D'_\phi$ is a diagonal matrix of phases, 
$c_{12}\equiv \cos\theta_{12}$, $s_{12}\equiv \sin\theta_{12}$ and
$\rho$ is the relative Majorana phase between the two mass eigenstates, 
varying between $0$ and $\pi$. With some easy algebra one finds 
$\sin2\theta=\sin\rho\sin2\theta_{12}$. Only this combination has physical 
meaning (it is measurable), while the mixing angle $\theta_{12}$ and the 
Majorana phase $\rho$ cannot be determined uniquely (a similar discussion 
can be found in \cite{JRS}).

The important consequence is that different small perturbations which 
generate $\Delta m^2_{12}$ may select very different values for $\theta_{12}$, 
which is the crucial parameter to determine the oscillation probability.
Let us consider a positive mass $\epsilon \ll a,b$. If $m_\nu$ diagonal 
entries are corrected as $a\rightarrow (a-\epsilon)$ and  
$-a\rightarrow (-a-\epsilon)$, then 
$\Delta m^2_{12} = 4\epsilon \sqrt{a^2+b^2}$ but the mixing angle is a free 
parameter, $\tan2\theta_{12}=b/a$, and the two neutrinos have opposite 
CP-parity, $\rho=\pi/2$. If instead $m_\nu$ off-diagonal entries are 
corrected as $b\rightarrow (b+i\epsilon)$, then $\Delta m^2_{12} = 4\epsilon a$ 
and the mixing angle is maximal, $\theta_{12}=\pi/4$, but the Majorana phase 
is a free parameter, $\sin\rho\approx b/\sqrt{a^2+b^2}$.\\


\noindent {\bf Quaternion model.~}
We now search for family symmetries that can lead to the structure in 
eq.(\ref{mag}). The equality $|m_{11}|=|m_{22}|$ cannot be explained by 
an Abelian symmetry, since in this case each lepton family would transform 
independently under the action of the symmetry group, so that equalities 
among independent mass matrix elements cannot be justified. Hence the 
two quasi-degenerate neutrino families should sit in a two-dimensional 
irreducible representation (2-dim irrep) of a non-Abelian group. Let us 
assign, therefore, the three Standard Model (SM) lepton doublets as follows:
\beq
\left(\bea{cc} L_1 \\ L_2 \eea\right) \sim {\bf 2} ~,~~~ L_3 \sim {\bf 1} ~,
\eeq
where the 1-dim irrep is not necessarily the singlet invariant under the 
symmetry group. To realize the structure in eq.(\ref{mag}), one needs two 
invariants, $a(L_1 L_1 -L_2 L_2)$ and $2b L_1 L_2$, and at the same time 
the combination $L_1 L_1 + L_2 L_2$ should not contribute.  We found that 
this $2\times 2$ pattern may be obtained by using the 2-dim irrep of any 
of the three smallest non-Abelian groups, $S_3$ \cite{s3}, $D_4$ \cite{d4} 
and $Q$. However, in the $S_3$ case it is not possible to maintain this 
pattern in a complete model with three families. In the $D_4$ case the 
combination $L_1L_1+L_2L_2$ is a group invariant and cannot be discarded 
without extra assumptions.

We therefore focus on the smallest quaternion group $Q$.  All the details on 
the group structure, the character table and our conventions for the irreps 
and their tensor products can be found in \cite{q8}.  For our purposes here 
it is sufficient to recall that $Q$ has four 1-dim irreps, ${\bf 1}^{++}$, 
${\bf 1}^{+-}$, ${\bf 1}^{-+}$ and ${\bf 1}^{--}$, with tensor product rules 
made obvious by the superscripts, and one 2-dim irrep ${\bf 2}$. The product 
of two $Q$-doublets $(\psi_1~\psi_2)^T$, $(\chi_1~\chi_2)^T\sim {\bf 2}$ 
goes as follows:
\beq
\bea{l}
(\psi_1\chi_2-\psi_2\chi_1)\sim {\bf 1^{++}} ~,~~~
(\psi_1\chi_1-\psi_2\chi_2)\sim {\bf 1^{+-}} ~,\\
(\psi_1\chi_2+\psi_2\chi_1)\sim {\bf 1^{-+}} ~,~~~
(\psi_1\chi_1+\psi_2\chi_2)\sim {\bf 1^{--}} ~.
\eea
\eeq

The charged lepton masses arise from the Yukawa coupling $y_{ijk}L_i e^c_j 
\phi_k$ where $\phi_k=(\phi^0,\phi^-)_k$ are Higgs doublets with vacuum 
expectation values (VEVs) $\langle\phi^0_k\rangle\equiv v_k$. The neutrino 
masses arise from the Majorana-type Yukawa coupling $f_{ijk}L_iL_j\Delta_k$ 
where $\Delta_k=(\Delta^{++},\Delta^+,\Delta^0)_k$ are Higgs triplets with 
$\langle\Delta^0_k \rangle = u_k$. These VEVs are naturally small when the 
triplets are super-heavy, by virtue of the type II seesaw mechanism.

Let us consider the following $Q$ assignments:
\beq\bea{l}
\left(\bea{cc} L_1 \\ L_2 \eea\right) \sim {\bf 2},~~ L_3 \sim {\bf 1}^{++} ,
~~ e^c_i \sim {\bf 1}^{-+},~{\bf 1}^{+-},~{\bf 1}^{--},\\
\left(\bea{c}\phi_1 \\ \phi_2 \eea \right)\sim {\bf 2},~~\phi_3\sim 
{\bf 1}^{--},~~ \Delta_i \sim {\bf 1}^{-+},~{\bf 1}^{+-},~{\bf 1}^{++} .
\eea\label{assign}\eeq
Then the charged lepton and neutrino mass matrices have the following 
structure:
\beq\bea{l}
m_l=\left(\bea{ccc} y_1v_2 & -y_2 v_1 & y_3 v_1 \\
y_1 v_1 & y_2 v_2 & y_3 v_2 \\
0 & 0 & y_4 v_3 \eea\right) ,\\
m_\nu = \left(\bea{ccc} f_2u_2 & f_1u_1 & 0 \\ f_1u_1 & -f_2u_2 & 0 \\ 0 & 0 
& f_3u_3 \eea\right) .
\label{mln}
\eea\eeq
The $1-2$ sector of $m_\nu$ has the desired form of eq.(\ref{mag}).
Therefore, in the CP conserving case, the $1-2$ mass degeneracy is realized.
Moreover, when $Q$ is broken by the Higgs doublet VEVs in the direction 
$v_1=0$, the unique off-diagonal element in the charged lepton mass matrix 
is $(m_l)_{23}$.  Therefore the $1-2$ mixing comes entirely from $m_\nu$, the 
$2-3$ mixing entirely from $m_l$ and one predicts $\theta_{13}=0$.
The scalar potential, the generation of VEVs and their alignment 
are discussed in the Appendix.

A possible extension of the $Q$ symmetry to the quark sector is discussed in 
\cite{q8}, where the phenomenological constraints on the extra Higgs doublets 
are also estimated. Notice that, since charged leptons mix only in the $2-3$ 
sector, there are no flavor-changing neutral currents (FCNCs) 
involving the electron, which are strongly constrained experimentally. 
Several possible tests of FCNCs in the $\mu-\tau$ sector
are discussed e.g. in \cite{CFM}.

Before performing a detailed analysis of lepton masses and mixing angles 
associated with the matrices in eq.(\ref{mln}) and their small perturbations, 
we would like to stress that the same pattern is maintained in many possible 
variants of the $Q$ model, with a different symmetry breaking sector, 
different field assignments and/or a different type of seesaw.

First, one may dislike the presence of multiple Higgs doublets at electroweak 
scale, because of sizable FCNC effects, a harsher hierarchy problem (more 
than one fine-tuning), worsened gauge coupling unification, etc.  Of course 
all such worries are based on some amount of theoretical prejudice. In any 
case, one can rephrase the flavor model above in terms of a unique Higgs 
doublet $\phi$ invariant under the family symmetry, then adding SM singlets 
$\varphi_i$ (flavons) charged under $Q$ as $({\bf 2},~{\bf 1}^{--})$. They 
enter charged lepton Yukawa couplings as $y_{ijk}L_i e^c_j (\varphi_k/\Lambda) 
\phi$ where $\Lambda > \langle \varphi_k \rangle$ is some cutoff scale. In 
this way one can reproduce the same mass matrix structure as before, while 
maintaining the SM particle content only at electroweak scale.  In this 
context, superheavy triplets may also be eliminated.  One may think of 
neutrino masses originating from the effective operator $f_{ijk} L_i L_j 
(\tilde{\varphi}_k/\Lambda) (\phi^* \phi^*/\Lambda)$, with flavons 
$\tilde{\varphi}_k$ charged under $Q$ as ${\bf 1}^{-+}$ and ${\bf 1}^{+-}$.

Second, a different structure of $m_l$ leading to the same mixing pattern as 
in eq.(\ref{mln}) can be obtained also when charged lepton singlets $e^c_i$ 
and lepton doublets $L_i$ both transform as (${\bf 2},~{\bf 1}^{++}$), which 
is required in left-right symmetric extensions of the SM. In this 
case, adding $\phi_4 \sim {\bf 1}^{+-}$, one finds
\beq
m_l=\left(\bea{ccc} y_3v_3 - y_4v_4 & 0 & -y_2 v_2 \\
0 & y_3 v_3+y_4 v_4 & y_2 v_1 \\
-y_1 v_2 & y_1v_1 & 0 \eea\right) ~.
\eeq
When $Q$ is broken by Higgs doublet VEVs in the direction $v_2=0$, one can 
accommodate $m_{e,\mu,\tau}$ and, at the same time, large (maximal) $2-3$ 
mixing.

Third, the neutrino mass matrix in eq.(\ref{mln}) can be derived with just a 
little bit more effort even in the context of type I seesaw, 
by introducing neutrino singlets $\nu^c_i\sim ({\bf 2},~{\bf 1}^{--})$. The 
neutrino Dirac mass matrix $m_D$ comes from the Yukawa coupling 
$y_{ij}L_i\nu^c_j \tilde{\phi}$, where $\tilde{\phi}=(\tilde{\phi}^+,
\tilde{\phi}_0)$ is a Higgs doublet transforming as ${\bf 1}^{--}$, 
so that $m_D =diag(x,x,y)$. The neutrino singlet mass matrix $M_R$ comes from 
the coupling $f_{ijk}\nu^c_i \nu^c_j S_k$, where $S_k$ are Higgs singlets with 
superheavy VEVs, transforming as $({\bf 1}^{-+},~{\bf 1}^{+-},
~{\bf 1}^{++})$. In this way one finds that $M_R$, $M_R^{-1}$ and 
$m_\nu\equiv -m_DM_R m^T_D$ all have the same structure as $m_\nu$ in 
eq.(\ref{mln}).  A similar model with type I seesaw which realizes the 
same form of $m_\nu$  by means of a $SU(2)\times U(1)$ family symmetry 
can be found in \cite{JRlast}. 
\\


\noindent {\bf Correlations among observables.~} 
The neutrino mass matrix 
in eq.(\ref{mln}) represents an interesting limit: $\theta_{13}=0$ and either 
$\Delta m^2_{12}=0$, in the case of no CP violation (that is, when 
$(f_2u_2)/(f_1u_1)$ is real), or $\theta_{12}=\pi/4$, when a 
nontrivial CP violating phase is present. 
In both cases solar oscillation data (see eq.(\ref{mix})) call for 
a perturbation to this matrix structure. In the CP-conserving case, the 
perturbation leads in general to a correlation between the values of 
$\Delta m^2_{12}$ and $\theta_{13}$. In the CP-violating case, the correlation 
will be between the deviation of $1-2$ mixing from maximal and nonzero $1-3$ 
mixing. Such correlations will be probed in future searches of $\theta_{13}$
\cite{glo}.

What may be the origin of such perturbations?  A first possibility is that 
the structures in eq.(\ref{mln}) are modified by radiative corrections from 
some large scale down to the electroweak scale. This scenario can be justified 
assuming that the family symmetry is broken at the large scale (by the VEVs 
of flavon fields) and that  $m_\nu$ runs below that scale as in the SM (or the 
MSSM). The radiative corrections to $m_\nu$ for matrix structures similar to 
those considered here have been studied in great detail in \cite{JRS,joshi,JM}.
It is found that it is possible to generate radiatively both $\Delta m^2_{12}$ 
and $\theta_{13}$, but it is problematic to obtain the present values of 
parameters in the Large Mixing Angle MSW region. We will not consider this 
radiative possibility in the following.

A more straightforward way to introduce a perturbation is to add extra Higgs 
multiplets, with different $Q$ assignments, which may provide a sub-dominant 
contribution to $m_\nu$. 
First, consider a Higgs triplet $\Delta_4\sim {\bf 1}^{--}$. 
Its VEV gives an equal contribution to the 
$11$- and $22$-entry of $m_\nu$. In the CP conserving (violating) case,
a small perturbation of this type generates 
$\Delta m^2_{12}\ne 0$ ($\theta_{12}\ne \pi/4$) 
but does not affect $\theta_{13}=0$.   Each 
observable is reproduced by a different parameter and therefore no 
correlations are predicted. In particular one cannot tell this scenario 
from any other model with $\theta_{13}=0$.

The most interesting scenario is obtained adding, instead, 
$(\Delta_4,~\Delta_5) \sim {\bf 2}$, with $u_4=0$ (same direction in group 
space as for the Higgs doublets $(\phi_1,~\phi_2)\sim {\bf 2}$ with $v_1=0$, see the Appendix). 
Then the neutrino mass matrix has the form
\beq
m_\nu = \left(\bea{ccc} a & b & d \\
b & -a & 0 \\
d & 0 & c \eea\right) ~,
\label{study}\eeq
where $d$ is proportional to $u_5$.  Let us recall that we are working in a 
basis where $m_l$ contains an arbitrary $2-3$ mixing $\theta_{23}^l$, 
so the $2-3$ mixing $\theta_{23}^\nu$ in 
$m_\nu$ is not required to match the observed value of 
$\theta_{23}=\theta_{23}^l+\theta_{23}^\nu$.
Neutrino mass matrices with one zero element and two independent nonzero 
elements equal to each other are a typical outcome of models based on the 
family symmetry $Q$ \cite{q8}. All possible matrices with this feature in 
the basis where $m_l$ is diagonal were analyzed in \cite{1+1}.

\noindent {\bf CP-conserving case.~} 
Let us diagonalize $m_\nu$ in eq.(\ref{study}) in  the case where all matrix 
elements are real  (without loss of generality one can take $a,~b$ and $d$ 
positive). In the limit $d\equiv \epsilon\ll a,b,|c|$, defining 
$m\equiv\sqrt{a^2+b^2}$ and expanding in $\epsilon$, the three neutrino 
masses are given by
\beq\bea{c}
m_1\approx m+\dfrac{\epsilon^2(m+a)}{2m(m-c)} ~,~~
m_2\approx -m+\dfrac{\epsilon^2(a-m)}{2m(m+c)} ~,\\
m_3\approx c +\dfrac{\epsilon^2(c+a)}{c^2-m^2} ~
\eea\eeq
(strictly speaking these equations hold only as long as 
$\epsilon\ll |m\pm c|$). The mass squared differences are then easily derived:
\beq\bea{l}
\Delta m^2_{12} \equiv m_2^2-m_1^2 \approx \dfrac{2m(a+c)}{c^2-m^2}\epsilon^2 
~,\\ \\
\Delta m^2_{23} \equiv m_3^2-m_2^2 \approx c^2 -m^2 ~.
\eea\eeq
The mixing angles are
\beq\bea{c}
\sin\theta_{13} \approx \dfrac{\epsilon(c+a)}{c^2-m^2} ~,~~~
\tan^2\theta_{12} \approx \dfrac{m-a}{m+a} ~,\\
\sin\theta^\nu_{23} \approx \dfrac{\epsilon b}{c^2-m^2} .
\label{mixe}\eea\eeq
The almost maximal $2-3$ mixing arises from the charged lepton sector. The 
ordering of the mass spectrum is normal (inverted) for $c^2-m^2 > 0$ ($<0$).
Since $\tan^2\theta_{12}<1$, solar neutrino data require $\Delta m^2_{12}>0$, 
that is $a+c>0$ ($<0$) for the case of normal (inverted) ordering. Finally 
and most importantly, the value of $\theta_{13}$ is correlated with the other 
observables:
\beq
\sin^2\theta_{13}\approx \frac 12 \frac{\Delta m^2_{12}}{\Delta m^2_{23}} 
\left(\cos2\theta_{12} + \frac{m_3}{m}\right) ~.
\label{corr}\eeq
In the case of normal ordering, $m_3>m$ and $\theta_{13}$ decreases by 
increasing the absolute neutrino mass scale, with a lower bound
\beq
\sin\theta_{13}|_{normal} > \cos\theta_{12} \sqrt{\Delta m^2_{12}/
\Delta m^2_{23}} \approx 0.15 ~.
\label{Nb}\eeq
In the case of inverted ordering $m_3 < -m\cos 2\theta_{12}$ and $\theta_{13}$ 
increases by increasing the absolute mass scale, with an upper bound
\beq
\sin\theta_{13}|_{inverted} < \sin\theta_{12} \sqrt{\Delta m^2_{12}/
|\Delta m^2_{23}|} \approx 0.10 ~.
\label{Ib}\eeq
These correlations are the main predictions of our model, which links the 
smallness of $\Delta m^2_{12}$ and $\theta_{13}$.

\begin{figure}[tbp]
\includegraphics[width=230pt]{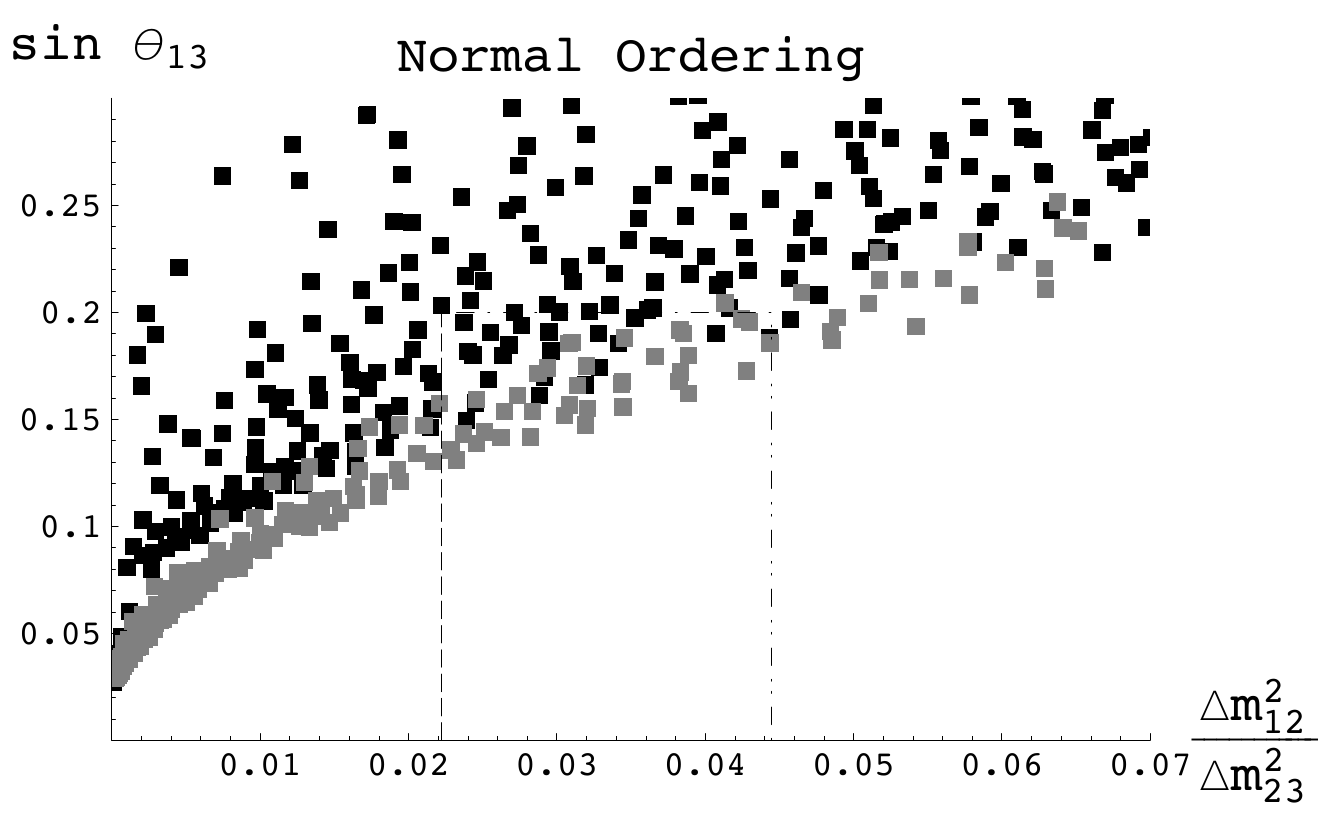}\\
\includegraphics[width=230pt]{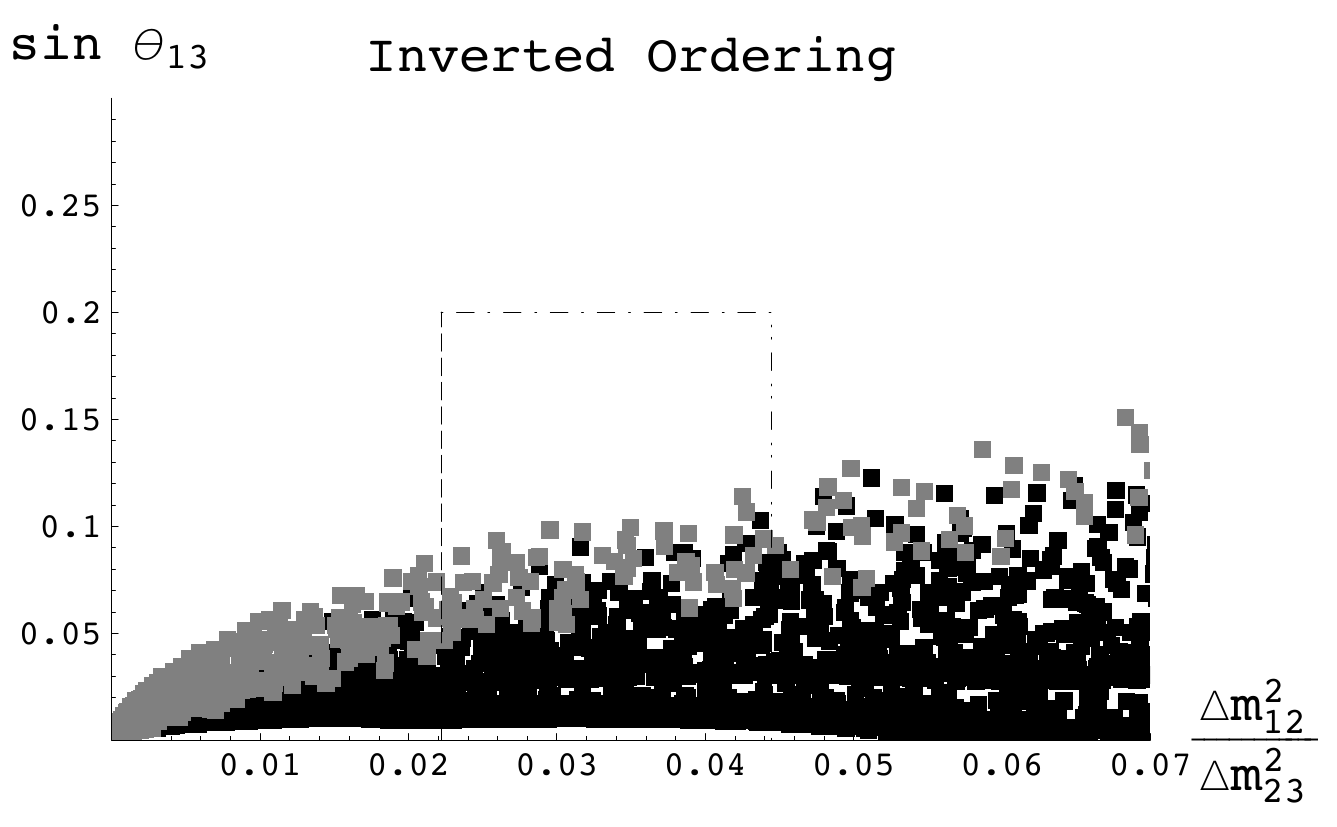}
\caption{\small The $1-3$ mixing as a function of $\Delta m^2_{12}/
\Delta m^2_{23}$ for the matrix $m_\nu$ in eq.(\ref{study}), assuming no CP 
violation. The upper (lower) panel corresponds to the case of normal 
(inverted) ordering of the mass spectrum. We imposed the constraints   
$0.24 <\sin^2 \theta_{12}<0.40$ ($3\sigma$ allowed interval)
as well as $\sum_i m_i < 1$ eV.
The dashed rectangle indicates the $3\sigma$ allowed region defined by
$0.022<\Delta m^2_{12}/\Delta m^2_{23}<0.045$ and $\sin\theta_{13}<0.2$.
The gray points are those where $d$ is significantly smaller (less than one 
third) than $a,~b$ and $|c|$.}
\label{s13vsR}
\end{figure}

\begin{figure}[tbp]
\includegraphics[width=230pt]{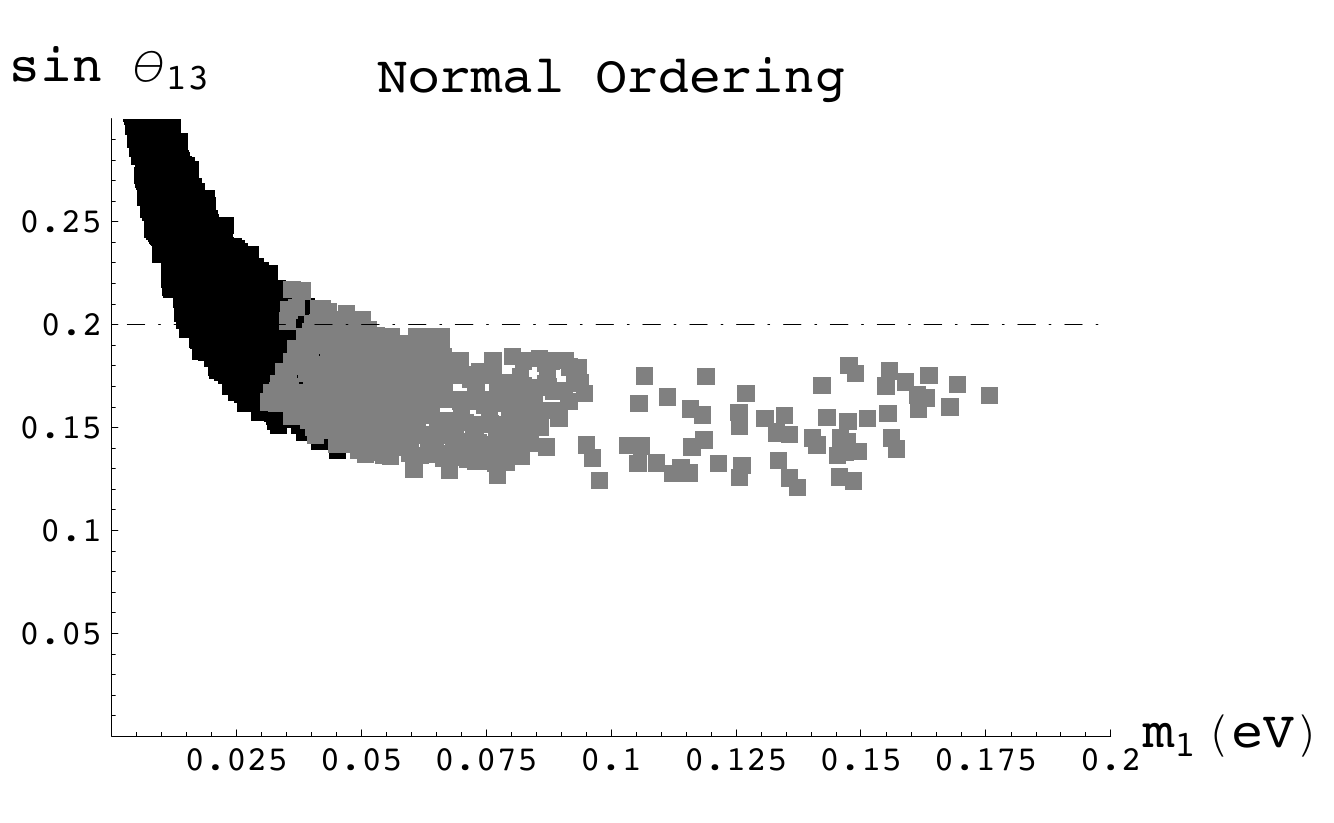}\\
\includegraphics[width=230pt]{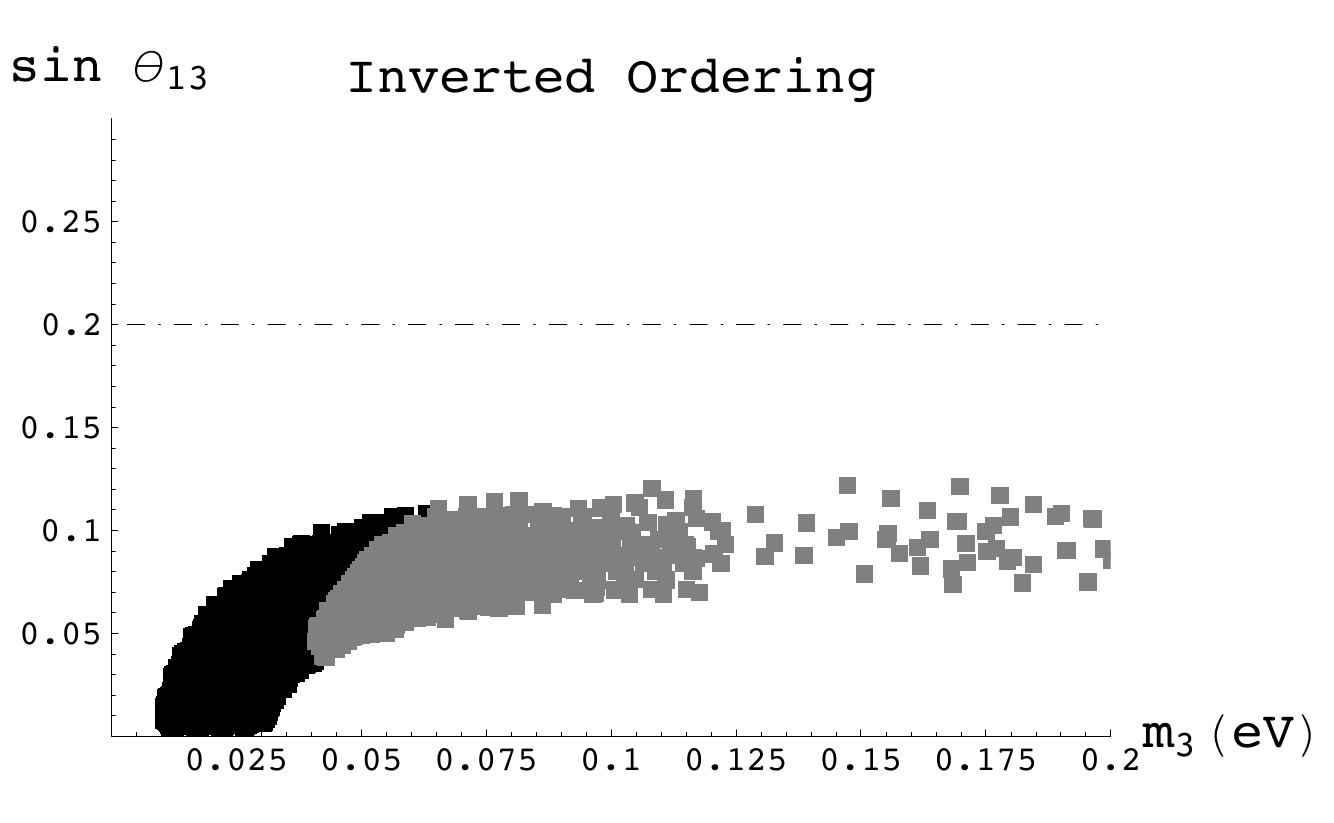}
\caption{\small The $1-3$ mixing as a function of the absolute mass scale
for the matrix $m_\nu$ in eq.(\ref{study}), assuming no CP violation.
The upper (lower) panel shows the allowed region in the $m_1 - \sin\theta_{13}$ 
($m_3 -\sin\theta_{13}$) plane for the case of normal (inverted) ordering of 
the mass spectrum. We imposed the constraints  
$0.022 < \Delta m_{12}^2/\Delta m_{23}^2<0.045$ and 
$0.24 <\sin^2 \theta_{12}<0.40$ ($3\sigma$ allowed intervals).
The dashed line indicates the $3\sigma$ upper bound $\sin\theta_{13}<0.2$.
The gray points are those where $d$ is significantly smaller (less than one 
third) than $a,~b$ and $|c|$.}
\label{NIreal}
\end{figure}

We performed a numerical analysis for the mass matrix in eq.(\ref{study}), 
scanning over the values of $a,~b,~c$ and $d$.
The prediction for $\sin\theta_{13}$ is
shown in Fig.\ref{s13vsR} as a function of the neutrino mass-squared-difference
ratio $\Delta m^2_{12}/\Delta m^2_{23}$.
The bounds (\ref{Nb}) and (\ref{Ib}), which assume best fit values of the 
measured parameters, are slightly relaxed but hold qualitatively:
we find $\sin\theta_{13}|_{normal} > 0.12$ and $\sin\theta_{13}|_{inverted} 
< 0.12$.

The prediction for $\sin\theta_{13}$ is
shown in Fig.\ref{NIreal} as a function of the absolute neutrino mass scale.
The lower density of 
allowed points for larger absolute mass scales indicates that a 
quasi-degenerate spectrum requires some fine-tuning in the input parameters 
$a,~b~,c$ and $d$. 
In the normal ordering case the experimental constraint $\sin\theta_{13}<0.2$ 
implies a lower bound $m_1\gtrsim 0.02$ eV. In the inverted ordering case 
$\sin\theta_{13}$ vanishes for $m_3 \sim 0.02$ eV. 
A lower bound on $\sin\theta_{13}$ holds also in the case of
inverted ordering if $m_3$ is sufficiently large. This can be probed
in neutrinoless $2\beta$ decay searches, which
can measure the effective mass parameter
$m_{ee} \equiv |m_{11}| =  a$. 
In the limit of small $d$ one 
has $m_{ee}\approx m\cos2\theta_{12}\sim m/2$. 
If $m_{ee}\gtrsim 0.04$ eV is found, our model predicts 
a lower bound $\sin\theta_{13}\gtrsim 0.05$. 

One should notice that the matrix in eq.(\ref{study}) may accommodate data 
even when $d$ is not much smaller that the other parameters, as shown by the 
black region in Figs.\ref{s13vsR} and \ref{NIreal}.  Indeed, $\Delta m^2_{12}$ 
and $\theta_{13}$ 
both vanish not only in the limit $d\rightarrow 0$, that was studied above, 
but also in the limit $c+a  \rightarrow  0$.  When $a+c=0$ one has
\beq\bea{c}
m_{1,2}=\pm m \equiv \pm \sqrt{a^2+b^2+d^2}~,~~~ m_3=-a~,\\
\tan \theta^\nu_{23}= - \dfrac{d}{b}~,~~~
\tan 2\theta_{12}=\dfrac{\sqrt{d^2+b^2}}{a} ~.
\eea\eeq
Therefore the ordering of the mass spectrum is inverted and both large angles 
can be accommodated (there is no need of large $2-3$ mixing in $m_l$). 
Defining $\epsilon \equiv a+c$ and taking the limit $|\epsilon| \ll a,b,d$, 
we find
\beq\bea{l}
\Delta m^2_{12} \approx \dfrac{2d^2 m\epsilon}{b^2+d^2} ~,\\
\sin\theta_{13} \approx \dfrac{bd\epsilon}{(b^2+d^2)^{3/2}} \approx
\dfrac 12 \dfrac{\Delta m^2_{12}}{\Delta m^2_{23}}\dfrac{\sin2\theta_{12}}
{\tan\theta^\nu_{23}} ~.
\eea\eeq
When $b\rightarrow 0$, $\theta_{13}$ vanishes but at the same time 
$\Delta m^2_{12}$ is nonzero.  This is why in the case of inverted ordering 
there is no lower bound on $\sin\theta_{13}$ (see lower panel in 
Figs.\ref{s13vsR} and \ref{NIreal}).

\noindent {\bf CP-violating case.~}
Let us consider the neutrino mass matrix  in eq.(\ref{study}) in the general 
case of complex matrix elements. 
In the limit $d=0$ one has $\theta_{12}=\pi/4$, $\theta_{13}=0$,
$\Delta m^2_{12}= 4{\rm Im}(ab^*)$ and $\Delta m^2_{23}= |c|^2-|a|^2-|b|^2$.
Notice that, in the presence of non-trivial phases, the smallness of 
$\Delta m^2_{12}/\Delta m^2_{23}$ is accidental.

When $d\equiv \epsilon$ is small, that is, $|\epsilon| \ll |a|,|b|,|c|$, 
a small  $\theta_{13}$ is generated and its value is correlated to 
the deviation from maximal $1-2$ mixing. 
One should diagonalize
\beq
m_\nu m_\nu^\dag =\left(\bea{ccc}
|a|^2+|b|^2+|\epsilon|^2 & ab^*-a^*b & a\epsilon^*+ c^*\epsilon \\
\dots & |a|^2+|b|^2 & \epsilon^* b \\
\dots & \dots & |c|^2 + |\epsilon|^2
\eea\right) ~.
\eeq
At leading order, we find
\beq\bea{l}
\sin\theta_{13}\approx \dfrac{|a\epsilon^* +  c^* \epsilon|}{\Delta m^2_{23}} ~,\\
\sin^2\theta_{12}\approx \dfrac 12
\left[1-\dfrac{|\epsilon|^2}{\Delta m^2_{12}}
\dfrac{|P_2(a,b,c)|}{\Delta m^2_{23}}\right] ~,
\eea\eeq 
where $P_2$ is a lengthy expression quadratic in $a,~b$ and $c$.
The form of 
$\sin\theta_{13}$ is analog to the one in eq.(\ref{mixe}).
The deviation from maximal $1-2$ mixing 
($\sin^2 \theta_{12}=1/2$) can be of order one  
for $|\epsilon|^2 \gtrsim \Delta m^2_{12}$.

\begin{figure}[tbp]
\includegraphics[width=230pt]{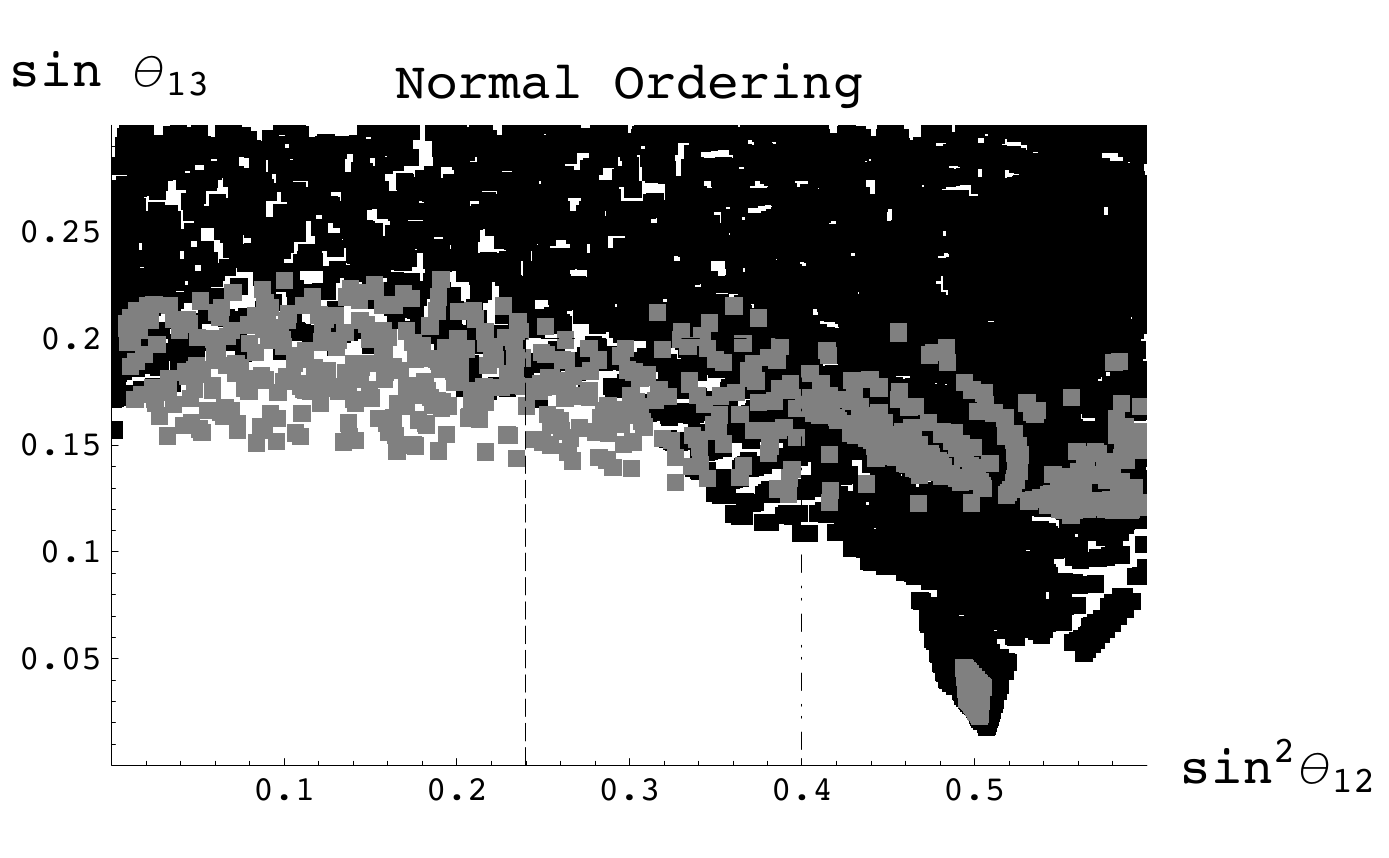}\\
\includegraphics[width=230pt]{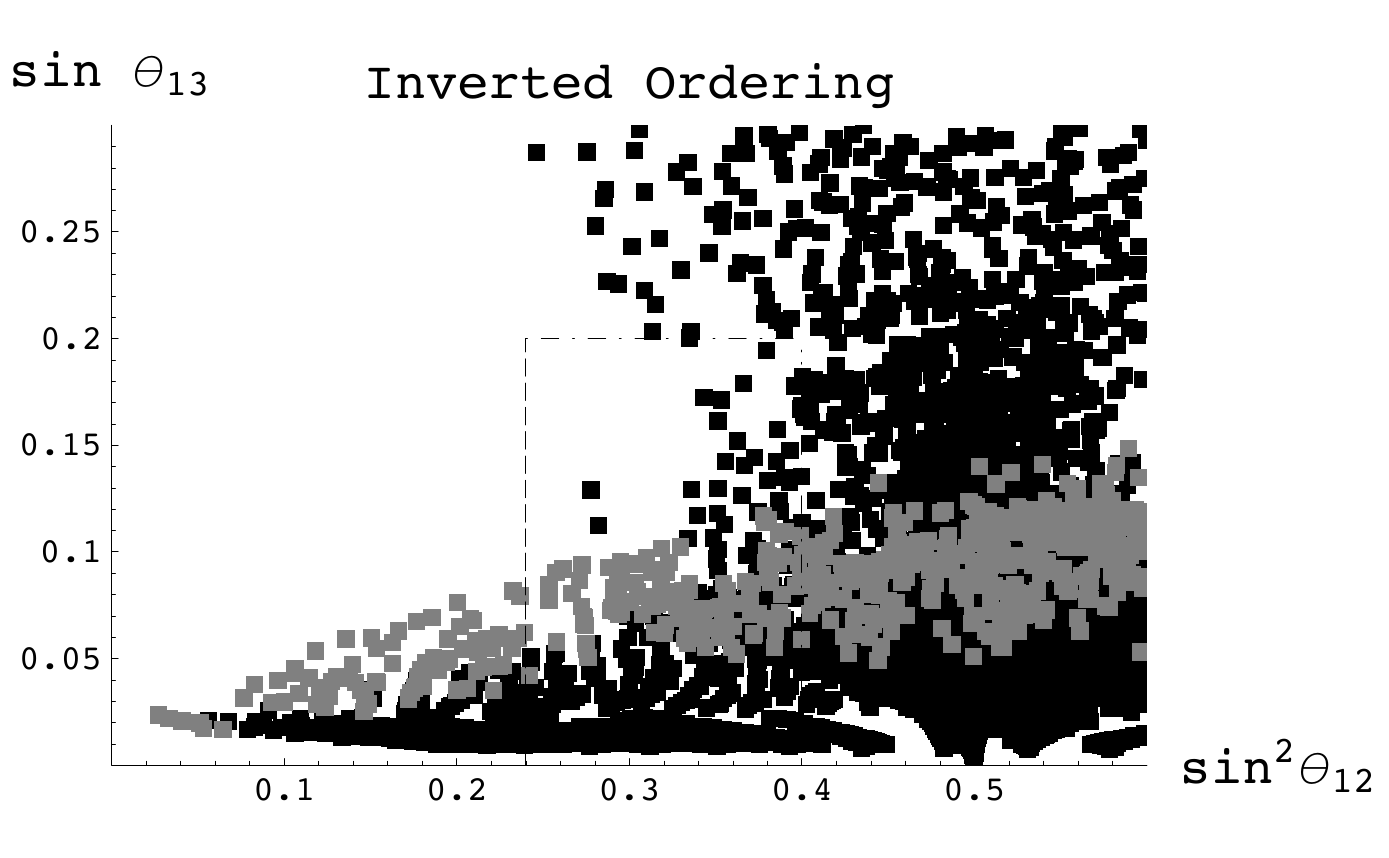}
\caption{\small The allowed region in the plane $\sin^2\theta_{12}-
\sin\theta_{13}$ for the matrix $m_\nu$ in eq.(\ref{study}) with the most 
general CP violating phases. The upper (lower) panel corresponds to the case 
of normal (inverted) ordering of the mass spectrum. We imposed the 
constraints $0.022 < \Delta m_{12}^2/\Delta m_{23}^2<0.045$ ($3\sigma$ 
allowed interval) as well as $\sum_i m_i < 1$ eV.
The dashed rectangle indicates the $3\sigma$ allowed region defined by
$0.24<\sin^2\theta_{12}<0.4$ and $\sin\theta_{13}<0.2$.
The gray points are those where $|d|$ is significantly smaller (less than one 
third) than $|a|,~|b|$ and $|c|$.}
\label{NIcomplex}
\end{figure}

We performed a numerical analysis for the most general choice of complex 
phases. The predicted correlation between $\sin^2\theta_{12}$ and 
$\sin\theta_{13}$ is shown in Fig.\ref{NIcomplex}.
We find that, in the normal ordering case, $\sin\theta_{13}\gtrsim 0.12$ is 
required to accommodate the non-maximal $1-2$ mixing.
Besides the opportunity to measure this value of $\theta_{13}$ already in the 
Double Chooz experiment, this scenario is also promising for future searches 
of leptonic CP violation \cite{glo}.
In the inverted ordering case, instead, there is no lower nor upper
bound on $\sin\theta_{13}$. However, when $\epsilon$ is small 
we find $0.05\lesssim \sin\theta_{13} \lesssim 0.12$ (gray region in the lower
panel of Fig.\ref{NIcomplex}). We checked that this is the case when 
$m_3\gtrsim 0.05$ eV, in analogy with 
the CP-conserving case (gray region in the lower panel of Fig.\ref{NIreal}).
\\


\noindent {\bf Conclusions.~} 
We studied the most general mass matrix for two mass-degenerate neutrinos. 
This is possibly a good limit to understand the smallness of the `solar' 
mass splitting $\Delta m^2_{12}$ relative to $\Delta m^2_{23}$. We have 
shown that such a mass degeneracy requires the equality of the $11$ and 
$22$ matrix elements, as well as the vanishing of one CP violating phase. 
The first requirement points to a non-Abelian family symmetry, the second 
indicates that CP violation in the lepton sector should be not generic, if present 
at all.  The matrix structure leading to the mass degeneracy is most easily 
accommodated, in the framework of three families, if  neutrinos only mix in 
the $1-2$ sector and, therefore, the large $2-3$ lepton mixing comes from 
the charged lepton sector.

We have shown that all these features can be explained by the simplest 
quaternion family symmetry $Q$, together with the requirement of no CP 
violation. We discussed several realizations of our $Q$ model, either 
employing several Higgs doublets or heavy flavon fields, and 
realizing the seesaw by either Higgs triplets or right-handed neutrinos. 
In the limit where $\Delta m^2_{12}=0$, the model predicts also 
$\theta_{13}=0$.

We discussed the possible perturbations generating nonzero $\Delta m^2_{12}$ 
and the consequent correlation with the nonzero value of $\theta_{13}$.  
We studied in detail the predictive case where $m_\nu$ depends only on three 
real parameters plus one small perturbation. Both normal and 
inverted ordering of the mass spectrum can be realized. In the normal case, 
$\sin\theta_{13}\gtrsim 0.12$ should be found, close to the present upper 
bound. In the inverted case, $\sin\theta_{13}$ is smaller than about 0.1. 
A lower bound ($\sin\theta_{13}\gtrsim 0.05$) holds in the inverted case
when the absolute neutrino mass scale $m_3$ (or equivalently 
the neutrinoless $2\beta$ effective mass $m_{ee}$) is larger than about 
$0.05$ eV.  

If CP violating phases are present, the $Q$ model predicts a correlation
between nonzero $\theta_{13}$ and the deviation of $\theta_{12}$ from 
the maximal value. Also in this scenario a lower bound 
$\sin\theta_{13}\gtrsim 0.12$ holds in the normal ordering case.\\

\noindent {\bf Acknowledgments.~} We thank the Aspen Center for Physics for 
hosting the 2007 workshop on ``Neutrino Physics: Looking Forward'' where we 
began our discussions.  MF also thanks the Department of Physics and 
Astronomy, University of California, Riverside, for hospitality during a 
subsequent visit. MF was supported in part by the CNRS/USA exchange grant 
3503 and by the RTN European Program MRTN-CT-2004-503369. EM was supported 
in part by the U.~S.~Department of Energy under Grant No.~DE-FG03-94ER40837.

\section*{Appendix: Alignment of VEVs}

The $Q$ model defined by eq.(\ref{assign}) contains three Higgs doublets and three
Higgs triplets. Here we will discuss some features of the scalar potential responsible for generating their VEVs. 

Let us consider the most general  $Q$-invariant potential 
for $(\phi_1,\phi_2)\sim{\bf 2}$ (notice that the conjugate $Q$-doublet is
$(\phi_2^*,-\phi_1^*)$):
\beq\bea{l}
V=m^2(\phi_1^\dag\phi_1+\phi_2^\dag\phi_2)\\ \\
+\frac 12 \lambda_1[(\phi_1^\dag\phi_1)^2+(\phi_2^\dag\phi_2)^2]
+\lambda_3(\phi_1^\dag\phi_1)(\phi_2^\dag\phi_2)\\ \\
+\frac 12 \lambda_2[(\phi_1^\dag\phi_2)^2+(\phi_2^\dag\phi_1)^2]
+\lambda_4(\phi_1^\dag\phi_2)(\phi_2^\dag\phi_1) ~.
\eea\eeq
It is bounded from below as long as
$\lambda_1 > |\lambda_2|+|\lambda_3+\lambda_4|$.
For $m^2<0$, the electroweak symmetry is broken 
in a minimum where $|v_1|^2=|v_2|^2=-m^2/(\lambda_1+\lambda_3+\lambda_4-|\lambda_2|)$.
The vacuum with $v_1=0$ and $v^2_2=-m^2/\lambda_1$ is a saddle.
In order to make it the minimum, one needs to break softly $Q$ to $Z_4$, by adding
\beq
V_{soft}=\mu^2 (\phi_2^\dag\phi_2-\phi_1^\dag\phi_1) ~.
\eeq
Notice that no other soft terms are allowed by the residual $Z_4$ symmetry.
Then the electroweak symmetry is broken if $m^2<|\mu|^2$ and the vacuum
with $v_1=0$ and $v_2^2=-(m^2+\mu^2)/\lambda_1$ is the absolute minimum
as long as $\mu^2(\lambda_1+\lambda_3+\lambda_4-|\lambda_2|) <
m^2(\lambda_1-\lambda_3-\lambda_4+|\lambda_2|)$.
This alignment of VEVs is the one analyzed in the text.

In fact, the analysis would be pretty much the same, even 
if one does not want to introduce $V_{soft}$. 
In this case the minimum is given by $v_1=v_2=v$ (strictly speaking the relative phase
between $v_1$ and $v_2$ can be $\pm1$ or $\pm i$, but for the sake of brevity we illustrate only one possibility). Then, the charged lepton mass matrix in eq.(\ref{mln})
can be rewritten as
\beq
m_l=\left(\bea{ccc}
\frac{1}{\sqrt{2}} & \frac{1}{\sqrt{2}} & 0 \\
-\frac{1}{\sqrt{2}} & \frac{1}{\sqrt{2}} & 0 \\
0 & 0 & 1 \eea\right)
\left(\bea{ccc} 0 & -\sqrt{2}y_2v  & 0 \\
\sqrt{2}y_1 v & 0 & \sqrt{2}y_3 v \\
0 & 0 & y_4 v_3 \eea\right) 
~.
\eeq
Therefore, up to a maximal $1-2$ rotation on the left and a harmless swap of 
$e^c_1$ and $e^c_2$ on the right, one recovers the same structure of $m_l$ as in the case $v_1=0$.
Notice that the maximal $1-2$ rotation does not modify the structure of the neutrino mass matrix $m_\nu$ in eq.(\ref{mln}), so that  all predictions for the lepton mixing angles are unchanged. However, such rotation introduces FCNCs involving the electron,
so that the non-standard Higgs bosons should be heavier to evade the experimental constraints. 

The addition of the third Higgs doublet $\phi_3\sim{\bf 1^{--}}$ does not change the qualitative features of the scalar potential discussed above.

Let us briefly discuss the origin of the VEVs of the three 
Higgs triplets in eq.(\ref{assign}), as well as of the two triplets
$(\Delta_4,\Delta_5)\sim{\bf 2}$, needed to generate $m_\nu$ in eq.(\ref{study}).
The $Q$-invariant scalar potential reads
\beq\bea{l}
V_\Delta= \sum_{i=1}^3 M^2_i \Delta_i^\dag \Delta_i
+ M^2_4 (\Delta_4^\dag \Delta_4 + \Delta_5^\dag \Delta_5) \\ \\
+\left[\mu_1 \Delta_1 \phi_1\phi_2 
+\mu_2 \Delta_2 (\phi_1\phi_1-\phi_2\phi_2) 
+\mu_3\Delta_3 \phi_3\phi_3 \right. \\ \\
\left.+\mu_4(\Delta_4\phi_1\phi_3+\Delta_5\phi_2\phi_3)+{\rm h.c.}\right]
+ V_\Delta^{quartic}~.
\eea\eeq
In the hypothesis that $M_i$ is much larger than the electroweak scale, one can integrate out $\Delta_i$ using its equation of motion. The triplet VEVs are thus determined as follows:
\beq\bea{c}
u_1^*=-\frac{\mu_1}{M^2_1} v_1 v_2 ~,~~~~
u_2^*=-\frac{\mu_2}{M^2_2} (v_1^2-v_2^2) ~,\\ 
u_3^*=-\frac{\mu_3}{M^2_3} v_3^2 ~,~~~~
u_{4,5}^*=-\frac{\mu_4}{M^2_4} v_{1,2} v_3 ~.
\eea\eeq
In the case $v_1=0$, one has automatically $u_4=0$ as assumed in eq.(\ref{study}), however also
$u_1$ vanishes, which is not acceptable. There are two ways to cure this problem.
The first is once again to add terms which break $Q$ softly, e.g. $V_\Delta^{soft} =
M_{12}^2 (\Delta_1^\dag\Delta_2+\Delta_2^\dag\Delta_1)$, that allows to induce both
$u_1$ and $u_2$ from $v_2$ only.

A second possibility without soft breaking is to resort to $Q$-invariant quartic couplings among Higgs doublets and triplets. Consider, in particular,
\beq
\lambda 
(\phi_1^\dag\phi_1-\phi_2^\dag \phi_2)(\Delta_3^\dag\Delta_1+\Delta_1^\dag\Delta_3)
\subset V_\Delta^{quartic} ~.
\eeq
Then the equation of motion for $\Delta_1$ leads to $u_1^* = \lambda u_3^* |v_2|^2 /M_1^2$. This contribution is tiny if all triplet masses are of the same order, but when
$M_1$ is instead close to the electroweak scale one can obtain $u_1\sim u_3$.

\bibliographystyle{unsrt}

\begin{thebibliography}{99}

\bibitem{MSTV}
  M.~Maltoni, T.~Schwetz, M.~A.~Tortola and J.~W.~F.~Valle,
  New J.\ Phys.\  {\bf 6} (2004) 122
  [hep-ph/0405172].
  
\bibitem{glo}
  D.~Duchesneau  [OPERA Collaboration],
  eConf {\bf C0209101} (2002) TH09
  [Nucl.\ Phys.\ Proc.\ Suppl.\  {\bf 123} (2003) 279]
  [hep-ex/0209082];
  F.~Ardellier {\it et al.}  [Double Chooz Collaboration],
  hep-ex/0606025;
  Y.~Itow {\it et al.}  [The T2K Collaboration],
  hep-ex/0106019;
  D.~S.~Ayres {\it et al.}  [NOvA Collaboration],
  hep-ex/0503053;
  P.~Huber, M.~Lindner, M.~Rolinec, T.~Schwetz and W.~Winter,
  Phys.\ Rev.\  D {\bf 70} (2004) 073014
  [hep-ph/0403068].
  
  

\bibitem{q8}
  M.~Frigerio, S.~Kaneko, E.~Ma and M.~Tanimoto,
  Phys.\ Rev.\  D {\bf 71} (2005) 011901
  [hep-ph/0409187];
%
  M.~Frigerio,
  hep-ph/0505144.
  
  
\bibitem{Getal}
  D.~Chang, W.~Y.~Keung and G.~Senjanovic,
  Phys.\ Rev.\  D {\bf 42} (1990) 1599;
  D.~Chang, W.~Y.~Keung, S.~Lipovaca and G.~Senjanovic,
  Phys.\ Rev.\ Lett.\  {\bf 67} (1991) 953.
  
\bibitem{Fetal}
  P.~H.~Frampton and T.~W.~Kephart,
  Int.\ J.\ Mod.\ Phys.\  A {\bf 10} (1995) 4689
  [hep-ph/9409330];
  P.~H.~Frampton and O.~C.~W.~Kong,
  Phys.\ Rev.\ Lett.\  {\bf 75} (1995) 781
  [hep-ph/9502395];
  P.~H.~Frampton and A.~Rasin,
  Phys.\ Lett.\  B {\bf 478} (2000) 424
  [hep-ph/9910522].
  
\bibitem{baku}
  K.~S.~Babu and J.~Kubo,
  Phys.\ Rev.\  D {\bf 71} (2005) 056006
  [hep-ph/0411226];
  Y.~Kajiyama, E.~Itou and J.~Kubo,
  Nucl.\ Phys.\  B {\bf 743} (2006) 74
  [hep-ph/0511268].
  
\bibitem{ACL}
  A.~Aranda, C.~D.~Carone and R.~F.~Lebed,
  Phys.\ Lett.\  B {\bf 474} (2000) 170
  [hep-ph/9910392];
  A.~Aranda, C.~D.~Carone and R.~F.~Lebed,
  Phys.\ Rev.\  D {\bf 62} (2000) 016009
  [hep-ph/0002044];
  A.~Aranda,
  0707.3661 [hep-ph].
  
\bibitem{Tp}
  P.~D.~Carr and P.~H.~Frampton,
  hep-ph/0701034;
  F.~Feruglio, C.~Hagedorn, Y.~Lin and L.~Merlo,
  Nucl.\ Phys.\  B {\bf 775} (2007) 120
  [hep-ph/0702194];
  M.~C.~Chen and K.~T.~Mahanthappa,
  0705.0714 [hep-ph];
  P.~H.~Frampton and T.~W.~Kephart,
  0706.1186 [hep-ph].
  
\bibitem{wolf}
  L.~Wolfenstein,
  Nucl.\ Phys.\  B {\bf 186} (1981) 147.
  
\bibitem{petcov}
  S.~T.~Petcov,
  Phys.\ Lett.\  B {\bf 110} (1982) 245.

\bibitem{JR}
  A.~S.~Joshipura and S.~D.~Rindani,
  Phys.\ Lett.\  B {\bf 494} (2000) 114
  [hep-ph/0007334].
  

\bibitem{JRS}
  A.~S.~Joshipura, S.~D.~Rindani and N.~N.~Singh,
  Nucl.\ Phys.\  B {\bf 660} (2003) 362
  [hep-ph/0211378].
  
\bibitem{s3}
An extensive list of references on $S_3$ flavor models can be found in the recent paper 
A.~Mondragon, M.~Mondragon and E.~Peinado,
  0706.0354 [hep-ph].
  
  
\bibitem{d4}
  W.~Grimus and L.~Lavoura,
  Phys.\ Lett.\  B {\bf 572} (2003) 189
  [hep-ph/0305046];
  W.~Grimus, A.~S.~Joshipura, S.~Kaneko, L.~Lavoura and M.~Tanimoto,
  JHEP {\bf 0407} (2004) 078
  [hep-ph/0407112];
  T.~Kobayashi, S.~Raby and R.~J.~Zhang,
  Nucl.\ Phys.\  B {\bf 704} (2005) 3
  [hep-ph/0409098];
  T.~Kobayashi, H.~P.~Nilles, F.~Ploger, S.~Raby and M.~Ratz,
  Nucl.\ Phys.\  B {\bf 768} (2007) 135
  [hep-ph/0611020].
  
\bibitem{CFM} 
  S.~L.~Chen, M.~Frigerio and E.~Ma,
  Phys.\ Lett.\  B {\bf 612} (2005) 29
  [hep-ph/0412018].


\bibitem{JRlast}
  A.~S.~Joshipura and S.~D.~Rindani,
  Phys.\ Rev.\  D {\bf 67} (2003) 073009
  [hep-ph/0211404].
  
\bibitem{joshi}
  A.~S.~Joshipura,
  Phys.\ Lett.\  B {\bf 543} (2002) 276
  [hep-ph/0205038].
  
\bibitem{JM}
  A.~S.~Joshipura and S.~Mohanty,
  Phys.\ Rev.\  D {\bf 67} (2003) 091302
  [hep-ph/0302181].
  
\bibitem{1+1}
  S.~Kaneko, H.~Sawanaka and M.~Tanimoto,
  JHEP {\bf 0508} (2005) 073
  [hep-ph/0504074].

\end{thebibliography}

\end{document}